\documentclass[pdflatex,sn-mathphys-num,iicol]{sn-jnl}

\usepackage{graphicx}%
\usepackage{multirow}%
\usepackage{amsmath,amssymb,amsfonts}%
\usepackage{amsthm}%
\usepackage{mathrsfs}%
\usepackage[title]{appendix}%
\usepackage{xcolor}%
\usepackage{textcomp}%
\usepackage{manyfoot}%
\usepackage{booktabs}%
\usepackage{algorithm}%
\usepackage{algorithmicx}%
\usepackage{algpseudocode}%
\usepackage{listings}%

\theoremstyle{thmstyleone}%
\theoremstyle{thmstyletwo}%

\theoremstyle{thmstylethree}%

\begin{document}

\title[]{Shape of liquid meniscus in open cells of varying geometry: a combined study via  simulation and experiment}


\author*[1]{\fnm{Konstantin S.} \sur{Kolegov}}\email{k.kolegov87@gmail.com}

\author[2]{\fnm{Viktor M.} \sur{Fliagin}}

\author[2]{\fnm{Natalia A.} \sur{Ivanova}}

\affil*[1]{\orgdiv{Mathematical Modeling Lab}, \orgname{Astrakhan Tatishchev State University}, \orgaddress{\street{20a~Tatishchev~str.}, \city{Astrakhan City}, \postcode{414056}, \state{Astrakhan Region}, \country{Russia}}}

\affil[2]{\orgdiv{Photonics and Microfluidics Lab, X-BIO Institute}, \orgname{ University of Tyumen}, \orgaddress{\street{6~Volodarsky~str.}, \city{Tyumen City}, \postcode{625003}, \state{Tyumen Region}, \country{Russia}}}


\abstract{Evaporative lithography in cells of arbitrary configuration allows for the creation of diverse particle deposition patterns due to the formation of a specific flow structure in the liquid caused by non-uniform evaporation. The latter in turn is determined by the shape of the liquid layer surface and the wetting menisci on the cell walls. Thus, predicting the shape of the wetting menisci can serve as a tool for controlling the process of creating desired particle deposition patterns and evaporation dynamics. Here, we propose a simple and sufficiently accurate methodology for determining the shape of the liquid meniscus in cells of arbitrary geometric shape, based on a combination of mathematical modeling and a series of experimental measurement techniques. The surface profiles of the liquid meniscus in cylindrical, square, and triangular cells were determined by measuring the change in the reflection angle of a laser beam from the free liquid surface while scanning from the cell wall to its center. The height of the wetting meniscus on the inner cell wall and the minimum liquid layer thickness at the center of the cell were measured by analyzing optical images and using a contact method, respectively. 3D meniscus profiles were obtained by numerically solving the Helmholtz equation. The boundary conditions and the unknown constant in the equation were determined based on experimental data obtained for several local points or cross-sections. The simulated meniscus shapes showed satisfactory agreement with the experimental local measurements, with a maximum relative error of less than 14\%.}

\maketitle

\section{Introduction}

The study of meniscus shapes (the ``liquid--air'' interface) in open liquid-filled cells is of interest for certain applications and technologies where self-assembly, evaporative lithography, and thermocapillary assembly can be employed~\cite{LebedevStepanov2013137,Kolegov2020,AlMuzaiqer2021126550,AlMuzaiqer2021,Ivanova2024}. These methods enable the production of geometrically structured deposits and coatings at macro-, micro-, and nanoscales. They can be applied in various fields: controlling virus transport in Petri dishes for biological studies~\cite{Lindsay2016}, protein detection in liquids~\cite{Banchelli2018}, manipulating the formation of 3D structures from colloidal particles for optoelectronic applications~\cite{Zhang2018}. Such menisci occur not only in cells but also in polygonal capillary tubes~\cite{Kialashaki2022}, microcavities of structured substrates~\cite{Ahn2015}, pixel wells~\cite{Wu2023}, etc. Liquid evaporation from open cells can occur under ambient conditions or in special regimes, such as reduced atmospheric pressure~\cite{Kazemi2018} or non-uniform cell heating~\cite{LebedevStepanov2013137,AlMuzaiqer2021126550,AlMuzaiqer2021,Ivanova2024}. Several methods exist for modeling meniscus shapes: energy minimization~\cite{Soligno2014,Devic2017,Soligno2024}, hydrostatic approximation~\cite{Bartashevich2009}, geometric approaches based on empirical formulas~\cite{Kialashaki2022}, etc.

Determining the liquid meniscus shape for different cell geometries is crucial because the curvature of the free liquid surface can affect the non-uniformity of the local vapor flux density $J(x,y,t)$. Over time $t$, during evaporation, the meniscus shape may become even more curved. Non-uniform evaporation along the free liquid surface induces flows that transport suspended and dissolved substances to specific regions, where the matter deposits at the cell bottom. Such systems can be used in evaporative lithography to produce various patterns of deposits and coatings. For further study of structured deposit formation processes in open cells of different geometries, a simple methodology for rapidly obtaining meniscus shape data is required.

The aim of the current work is to develop a simple and effective methodology for obtaining data on meniscus shapes in cells with different geometries.

\section{Methods}

\subsection{Methodology}
\begin{figure}[h]
	\centering
	\includegraphics[width=0.4\textwidth]{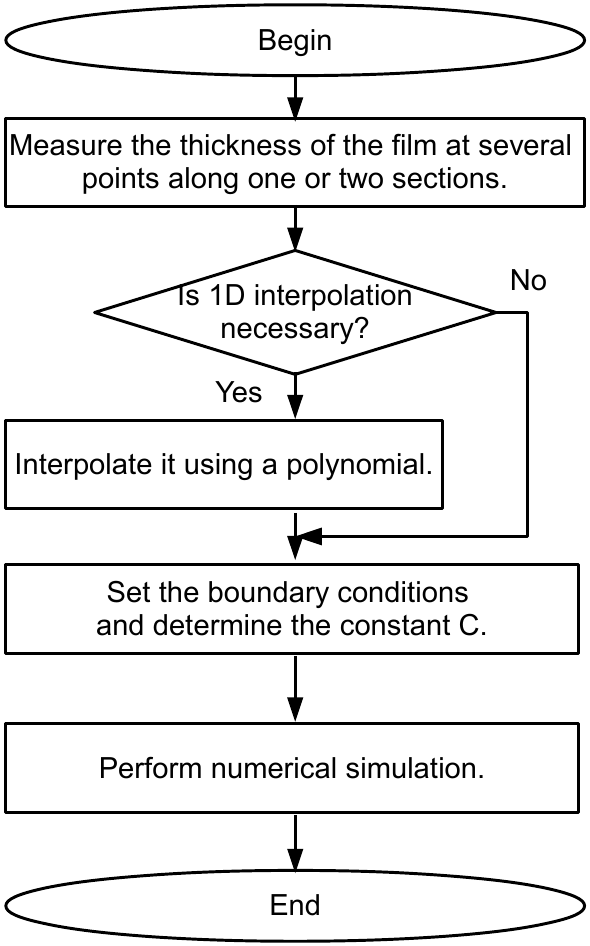}
		\caption{The proposed methodology presented as an algorithmic flowchart.}
	\label{fig:methodology}
\end{figure}

The methodology proposed here for obtaining data on the meniscus shape in a liquid-filled cell is shown in Fig.~\ref{fig:methodology}. Each individual step is described in detail in the corresponding subsections below.

\subsection{Experiment}

\subsubsection{Fabrication of cells and experimental preparation}

Experiments were performed in cells with borders shaped as equilateral triangles, squares, and circles with equal internal area $S =$ 144~mm$^2$. The borders in the form of selected shape contours were cut using a laser cutter (Halk~90, 60~W, precision 0.02~mm, China) from a 3.28~mm thick PMMA sheet and glued to the surface of dark glass, thus forming the cell. The side length of the square cell was $12.14\pm 0.17$ mm, the triangular cell~--- $18.11\pm 0.38$ mm, and the diameter of the cylindrical cell was $13.47\pm 0.15$ mm. The fabricated cells were washed with isopropyl alcohol and distilled water, then dried under an air stream. The wettability of the border was evaluated by the contact angle $\theta$ of a distilled water droplet (GFL 2002 distiller, 2.3 $\mu$S/cm at 25$^\circ$C, Germany) with a volume of approximately 3 $\mu$L, using a contour analysis system (OCA 15, DataPhysics Instruments, Germany). The measured water contact angle value for the border was $\theta = 76.6 \pm 0.3^\circ$. The cells were filled with 300~$\mu$L of distilled water using a 100~$\mu$L pipette (Thermo Fisher Scientific, Russia, accuracy 1.5\%), by adding three successive aliquots of 100 $\mu$L each. The volume setting error was $\pm$5 $\mu$L. After which, measurements of the liquid surface profiles in the cell and the wetting meniscus profiles on the borders were performed.

\subsubsection{Method of measuring wetting meniscus profiles}

\begin{figure*}
	\centering
	\includegraphics[width=0.75\textwidth]{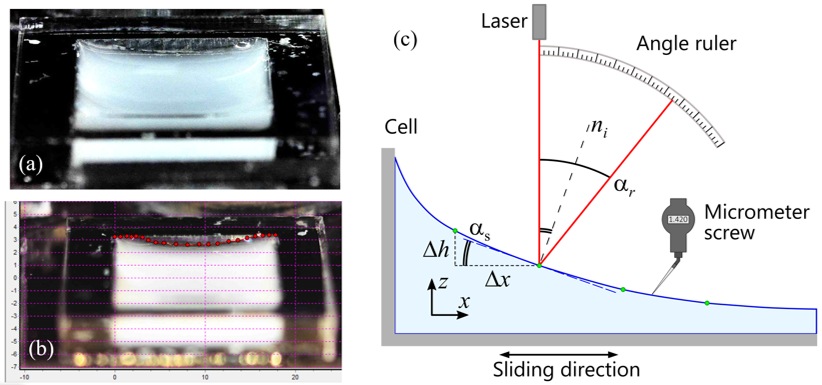}
	\caption{(a) Picture of the liquid wetting meniscus on the inner border of a square cell; (b) Example of picture processing in Graph2Digit software, red points~--- coordinates of the three-phase contact line; (c) Schematic of liquid surface inclination measurement using a laser beam and layer thickness measurement using a micrometric probe (contact method).}
	\label{fig:MeniscusMeasureExperimentSetup}
\end{figure*}

The wetting meniscus profile on the inner cell wall was measured as follows. Using a camera (Logitech Brio with Kowa HR F2/35 lens), an image of the inner wall of an empty cell was captured to establish the scale. Then, the cell was filled with liquid and an image of the wall with the meniscus was captured, Fig.~\ref{fig:MeniscusMeasureExperimentSetup}(a), using the same camera settings. Next, in the Graph2Digit software, using a coordinate grid, the coordinates of points along the wetting meniscus boundary on the cell wall were determined, Fig.~\ref{fig:MeniscusMeasureExperimentSetup}(b). To enhance image contrast, the cell border was backlit with an LED lamp from the opposite side.

The vertical coordinate of the liquid surface (liquid layer thickness) at the cell center was measured using a contact method with a Sylvac $\mu$s246 digital micrometer (instrument error $\pm5$ $\mu$m, Switzerland) equipped with a probe (a pipette tip); see  Fig.~\ref{fig:MeniscusMeasureExperimentSetup}(c). The zero reference corresponded to the micrometer reading when the probe touched the bottom of the empty cell. The probe was then raised, the cell was filled with liquid, and the probe was lowered again until it contacted the liquid surface. The vertical coordinate of the liquid surface was determined from the difference in micrometer readings.

The cross-sectional profile of the surface from the cell center to its wall was determined by measuring changes in the reflection angle of a laser beam from the free liquid surface. A semiconductor laser beam was directed vertically downward onto the substrate and focused using a built-in lens so that the beam caustic was positioned approximately 1 mm above the cell bottom. The beam reflected from the free liquid surface hit an arced screen with angular markings, which was used to determine the surface inclination angle, $\alpha_s = \alpha_r/2$, Fig.~\ref{fig:MeniscusMeasureExperimentSetup}(c). The distance from the reflection point on an undeformed layer to the screen was 250~mm, allowing small changes (up to 2~mm) in the reflection point height due to the meniscus to be neglected.

The surface inclination angle measurements started near the cell wall, as this provided a well-defined initial horizontal coordinate. The cell was then translated relative to the stationary laser beam in 0.5 mm steps from the wall toward the center. However, because the laser beam has a finite size, the image of the reflected beam on the screen was smeared due to the influence of the wall. For this reason, we moved away from the wall by one scanning step (0.5 mm) to obtain the first experimental data point. This makes it problematic to measure the meniscus shape near the cell wall with high detail. The resulting array of surface inclination points was interpolated using a third-order spline function, and the surface profile function was then obtained by integration. The algorithm is described in detail in Ref.~\cite{Klyuev2021}. The resulting surface profile function was referenced to the measured coordinates of the wetting meniscus height on the inner cell wall and the minimum layer thickness at the cell center.

\subsection{Simulation}
\subsubsection{Model equations}
For capillary numbers $\mathrm{Ca} = \eta v_c /\sigma \ll 1$, the hydrostatic approximation can be used ($\eta$ is the dynamic viscosity, $v_c$ is the characteristic flow velocity, and $\sigma$ is the surface tension coefficient)~\cite{Bartashevich2009}. Let's take the values of these parameters for water at room temperature: viscosity $\eta \approx 10^{-3}$ sPa, capillary flow velocity $v_c \approx$ 10~$\mu$m/s (for slow evaporation~\cite{Hamamoto2011}), and surface tension $\sigma \approx$ $73\cdot 10^{-3}$~N/m. Thus, we obtain a capillary number value $\mathrm{Ca} \approx 10^{-7}$, so we will use the hydrostatic approximation. The pressure under the curved meniscus is given by $$P=P_0 + \sigma \kappa,$$ where $P_0$ is the atmospheric pressure and $\kappa$ is the mean surface curvature,
\begin{equation}\label{eq:SurfaceCurvatureFormula}
	\kappa = \frac{2h_x h_y h_{xy} - h_{xx}(1+h_y^2) - h_{yy}(1+h_x^2)}{(1+h_x^2+h_y^2)^{3/2}},
\end{equation}
where $h(x,y)$ is the local liquid layer height~\cite{Barash2009}. Here the subscripts $x$ and $y$ denote partial derivatives with respect to the corresponding coordinates. Let's write the equilibrium condition taking into account the hydrostatic pressure at the bottom of the cell, $$P - P_0 + \rho g h = \mathrm{C},$$ $$\sigma \kappa + \rho g h = \mathrm{C},$$ where $\mathrm{C}$ is the unknown constant, $g$ is the gravitational acceleration ($g \approx$ 10~m/s$^2$), and $\rho$ is the liquid density ($\rho \approx$ $10^3$~kg/m$^3$). Knowing the volume of the liquid, one can estimate the Bond number, $\mathrm{Bo} = L^2 g \rho / \sigma \approx 6$. This moderate value confirms that in this problem, it is necessary to account for both gravity and surface tension simultaneously. Here $L$ denotes the characteristic size, $L = \sqrt[3]{V} \approx$ $6.7\times 10^{-3}$~m. If we assume that the two-phase boundary is weakly curved, then formula~\eqref{eq:SurfaceCurvatureFormula} simplifies~\cite{LandauLifshitz2001},
 \begin{equation}\label{eq:SurfaceCurvatureSimplified}
 	\kappa = - (h_{xx} + h_{yy}).
 \end{equation}
Under the assumption of a weakly curved interface, which justifies the use of Eq.~\eqref{eq:SurfaceCurvatureSimplified}, the governing equation reduces to the Helmholtz equation for the meniscus shape in the cell~\cite{LandauLifshitz2001}
\begin{equation}\label{eq:HelmholtzEquation}
-\sigma (h_{xx} + h_{yy}) + \rho g h = \mathrm{C}.
\end{equation}

\subsubsection{Model applicability limits}

Any model has its limits of applicability, and our model is no exception. Let us discuss the cases where using this model is permissible.

First, the determining factor is the thickness of the liquid layer in question, even approximately. If the layer thickness is 100 nm or even less, the disjoining pressure becomes the dominant factor, which is not accounted for here. It results from the combined action of molecular forces, which begin to play a decisive role at small distances. Thus, it is unacceptable to describe the shape of an ultrathin film using this model.

Second, the hydrostatic approximation is only valid for the case of $\mathrm{Ca} \to 0$. A significant increase in liquid viscosity or velocity of capillary flow, along with a decrease in surface tension, can lead to the model being unsuitable for describing the meniscus shape. This depends on the physical properties of the specific liquid. If dealing with a solution, it also depends on its concentration, including the presence of surfactants. The capillary flow, for instance, can be influenced by evaporation intensity. Evaporation increases with decreasing atmospheric pressure, increasing temperature, and decreasing air humidity; furthermore, it depends on the physical properties of the liquid. Thus, the case of relatively fast evaporation may fall outside the model's applicability limits. For capillary number $\mathrm{Ca} \geq 1$, the influence of viscous stresses becomes significant, and the hydrostatic approximation underlying this model is no longer valid.

Third, it is necessary to understand in which cases the film can be considered weakly curved, allowing the use of the simplified formula~\eqref{eq:SurfaceCurvatureSimplified} instead of~\eqref{eq:SurfaceCurvatureFormula}. The main requirement can be formulated mathematically as $h_{xx} + h_{yy} \ll 1$. When this condition is met, the film can be considered weakly curved. Here, we are close to exceeding the model's applicability limits when considering the film wetting the cell wall. For this reason, it is highly desirable in the future to transition to using formula~\eqref{eq:SurfaceCurvatureFormula}. Here, however, formula~\eqref{eq:SurfaceCurvatureSimplified} is used to demonstrate the proposed methodology for obtaining meniscus shape data with a simple example.

\section{Results and discussions}

\begin{figure*}[h]
    \begin{minipage}[t]{0.49\textwidth}
        \centering
        \includegraphics[width=0.99\textwidth]{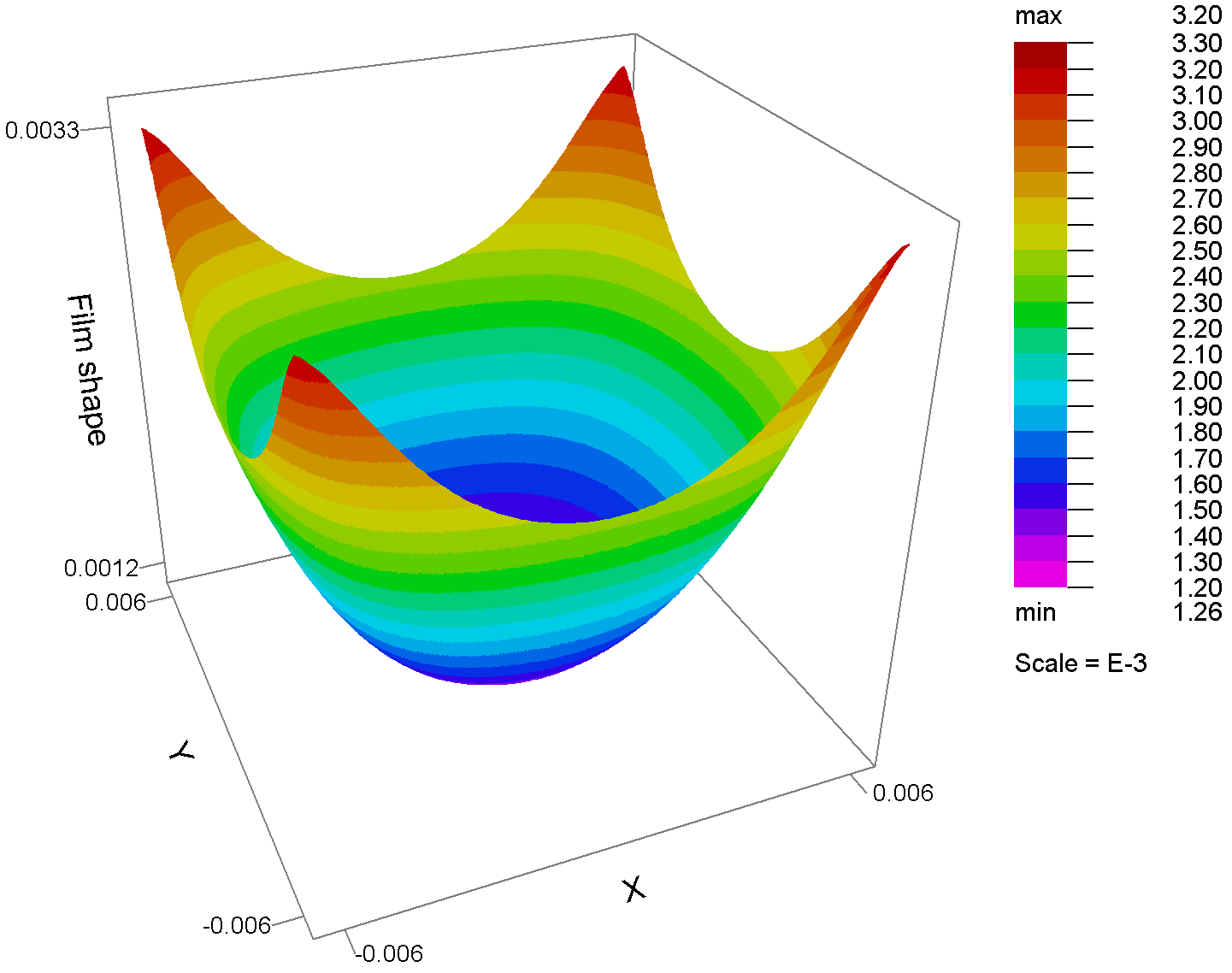}\\(a)
    \end{minipage}
    \begin{minipage}[t]{0.49\textwidth}
        \centering
        \includegraphics[width=0.99\textwidth]{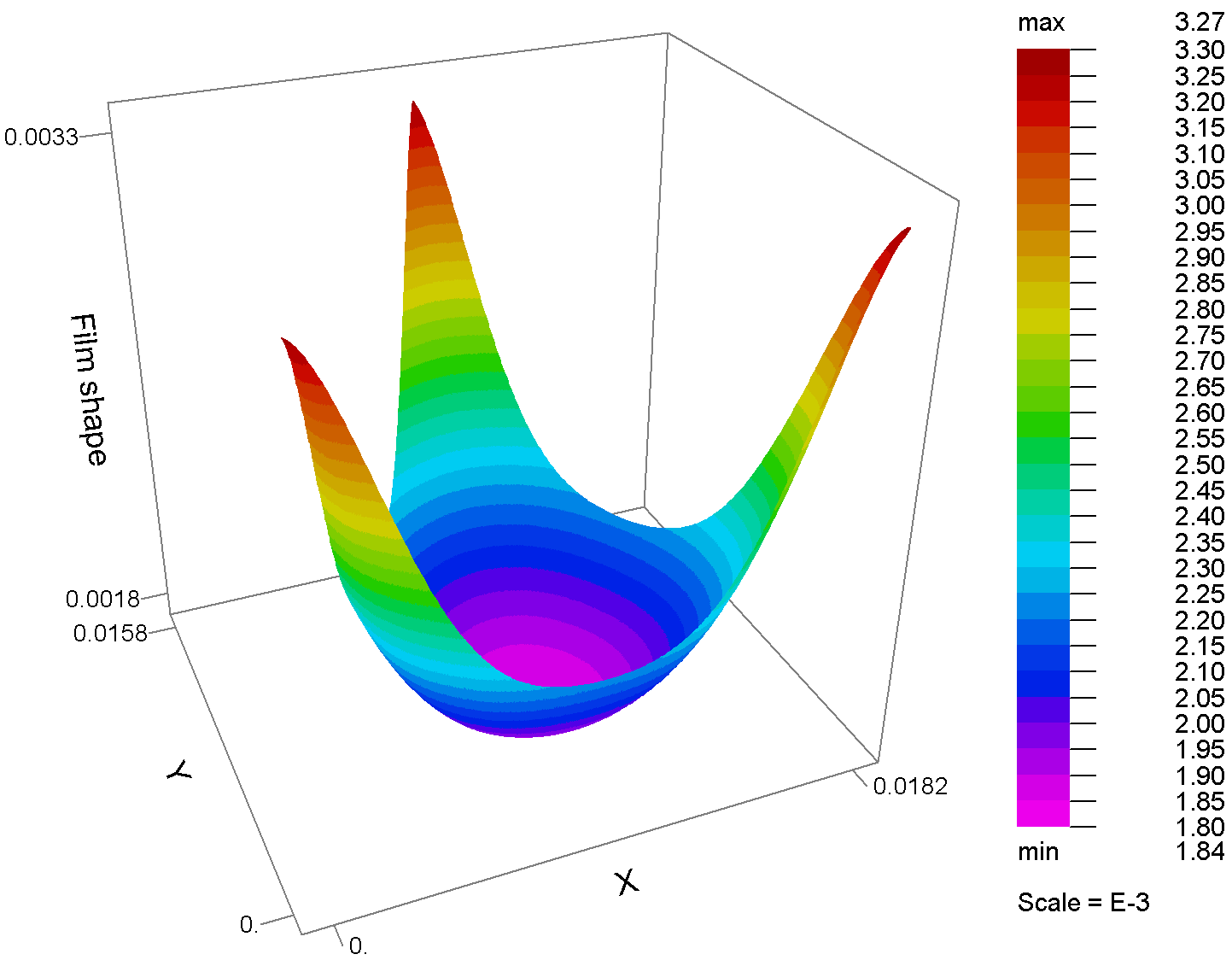}\\(b)
    \end{minipage}
	\caption{Results of numerical simulation of the meniscus shape in (a) square and (b) triangular cells.}
	\label{fig:filmShape3D}
\end{figure*}

The numerical solution of equation~\eqref{eq:HelmholtzEquation} depends on the boundary conditions and the value of the constant $\mathrm{C}$. These details are described comprehensively in appendices~\ref{app:SquareCell}, \ref{app:TriangularCell}, and \ref{app:CylindricalCell} for different cell geometries. Figure~\ref{fig:filmShape3D} shows 3D profiles of the menisci for square and triangular cells (calculation performed using the finite element method in FlexPDE 7.21 Professional). Due to the capillary effect, the maximum liquid film height occurs at the cell corners. The minimum liquid layer height is located in the central part of the cells. Thus, the meniscus surface profiles are concave, which is related to the hydrophilic properties of the cell walls. In the square cell, the maximum film height variation is approximately 2 mm. For the triangular cell, this value is about 1.4 mm.


\begin{figure}[h]
	\centering
	\includegraphics[width=0.35\textwidth]{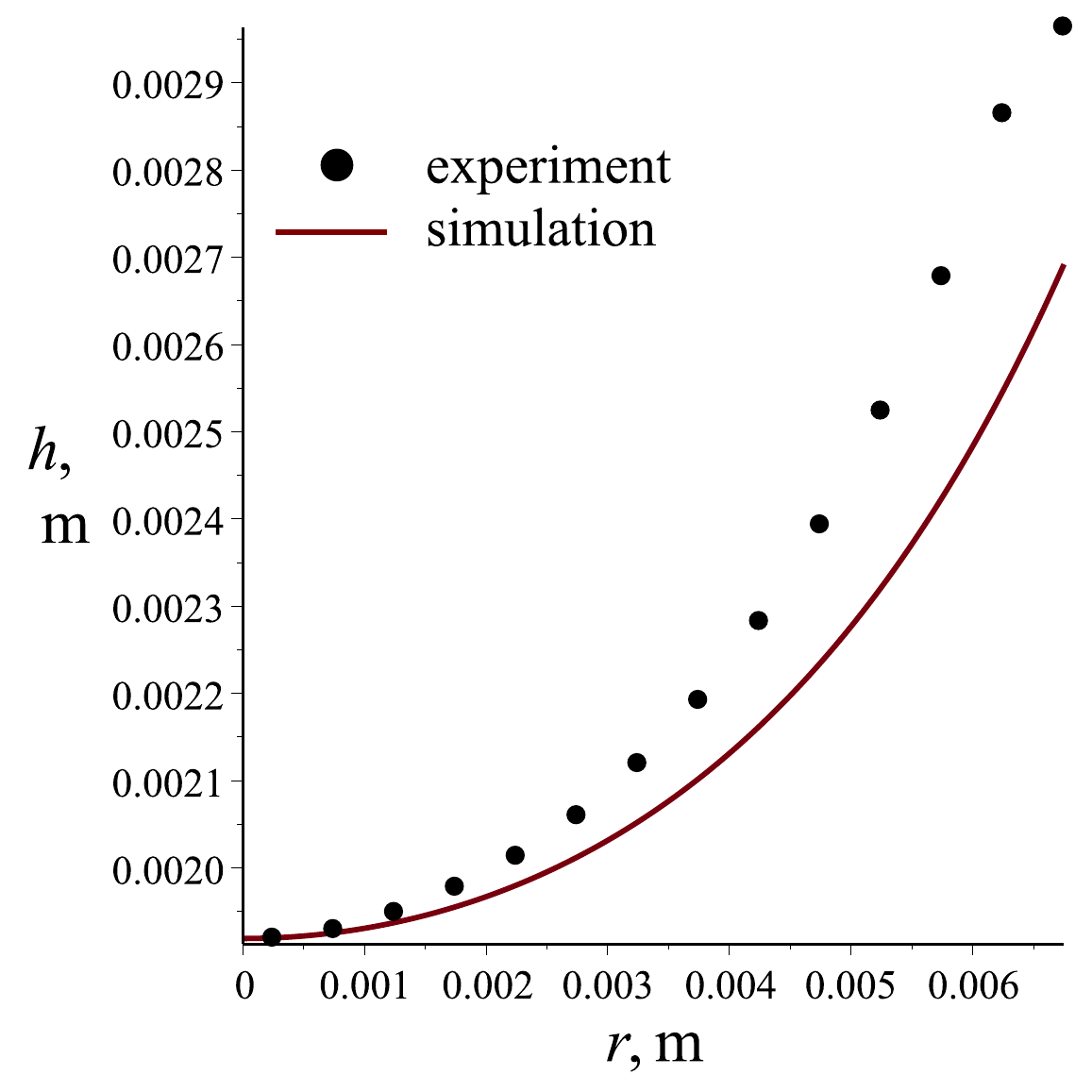}
	\caption{Results of numerical simulation and experimental measurements of the meniscus shape in a cylindrical cell.}
	\label{fig:h_experVsNumericCylinder}
\end{figure}

Considering the axial symmetry in cylindrical cells, the problem can be simplified by considering the dependence $z = h(r)$, where $r$ is the radial coordinate. The simulation results of the free liquid surface shape in a cylindrical cell (using the dsolve module in Maple~18) show good qualitative agreement with experimental data (Fig.~\ref{fig:h_experVsNumericCylinder}). We observe that near the cell wall, the liquid layer rises due to wetting of the solid surface. The thickness variation of the liquid layer is approximately 1 mm according to experimental data. In the central part of the cell, the calculated and experimental results are close in value. However, as $r$ increases, a discrepancy between the results becomes apparent. Near the cell wall, the numerically obtained value is $h(r=R) \approx$ 2.7~mm, while the experimental value corresponds to $h(r=R) \approx$ 2.96~mm (the cylindrical cell radius $R\approx$ 6.55~mm). Thus, the simulation results predict the meniscus height at the cell wall to be 8.8\% lower than observed experimentally. In this case, this value represents the maximum simulation error.

\begin{figure}[h]
	\centering	\includegraphics[width=0.4\textwidth]{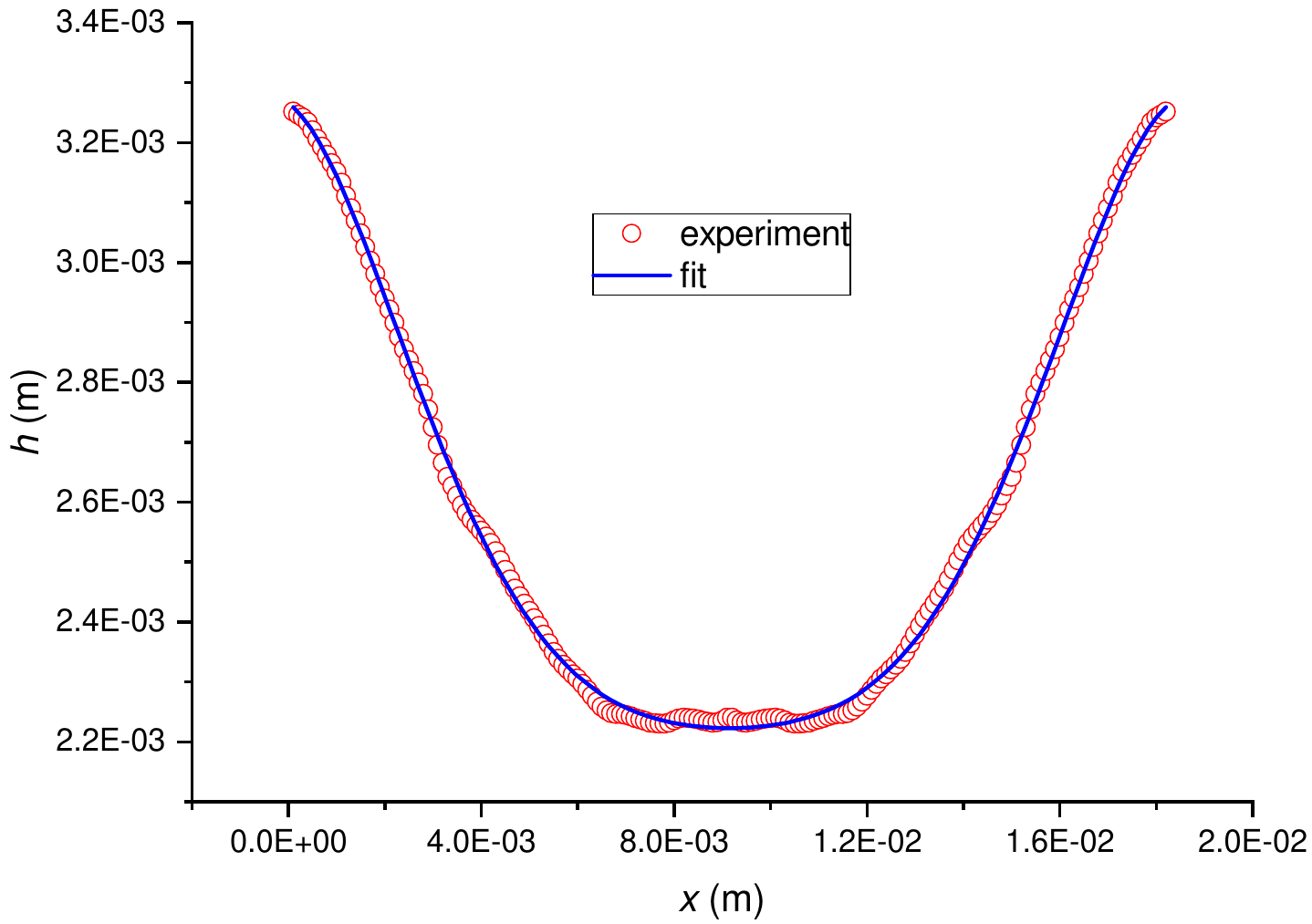}(a)\\ \includegraphics[width=0.4\textwidth]{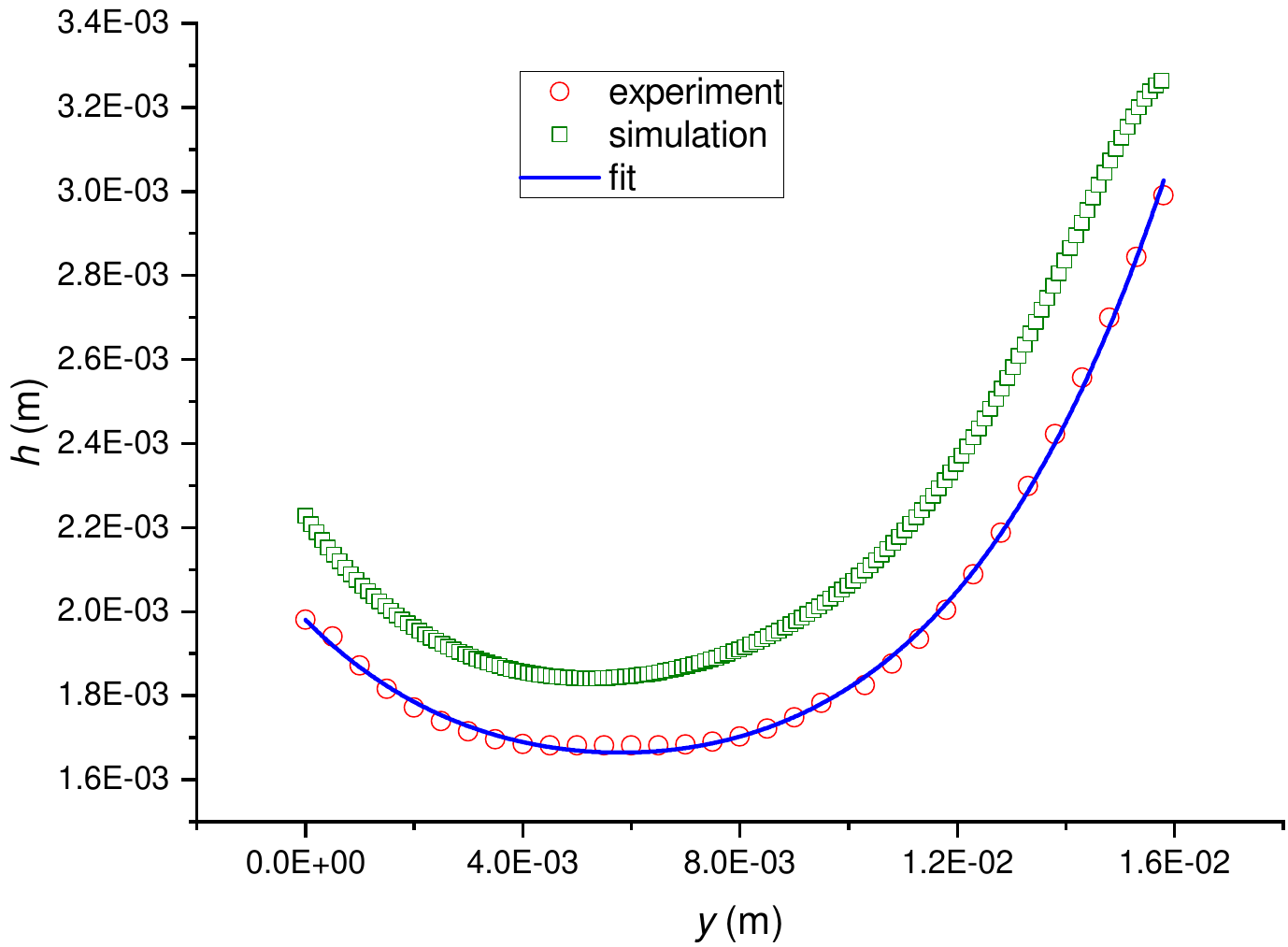}(b)
	\caption{Shape of the two-phase interface (``liquid--air'') (a) along the wall of triangular cell measured by images of the meniscus on the wall, and (b) along its median measured by the laser scanning method.}
	\label{fig:filmShapeInTriangularCell2D}
\end{figure}

The measurement results of the two-phase interface (``liquid--air'') shape along the wall of a triangular cell and along its median connecting the triangle's corner with the midpoint of the opposite wall are shown in Fig.~\ref{fig:filmShapeInTriangularCell2D}. Approximation of experimental data for the liquid film thickness along the cell wall (Fig.~\ref{fig:filmShapeInTriangularCell2D}a) was used as a boundary condition for numerical calculations (see details in Appendix~\ref{app:TriangularCell}). The simulation results show a reasonable agreement with the experimental trends, though with a quantifiable deviation (Fig.~\ref{fig:filmShapeInTriangularCell2D}b). The model prediction of liquid layer thickness in the triangular cell is slightly overestimated compared to experimental data. The maximum error along this cross-section is observed at point $y \approx$ 14~mm and amounts to approximately 13\%.

\begin{figure}[h!]
	\centering	\includegraphics[width=0.4\textwidth]{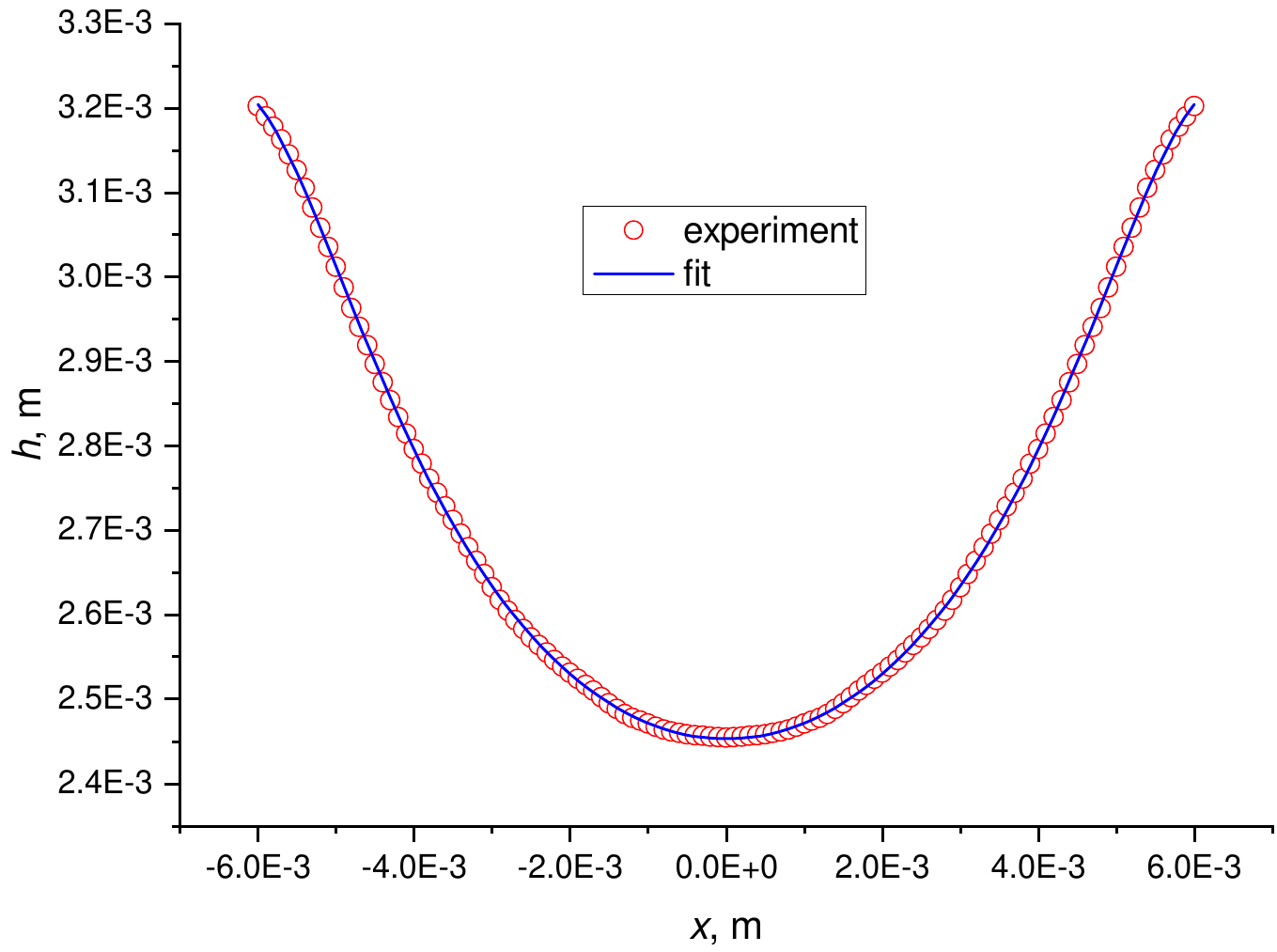}(a)\\ \includegraphics[width=0.4\textwidth]{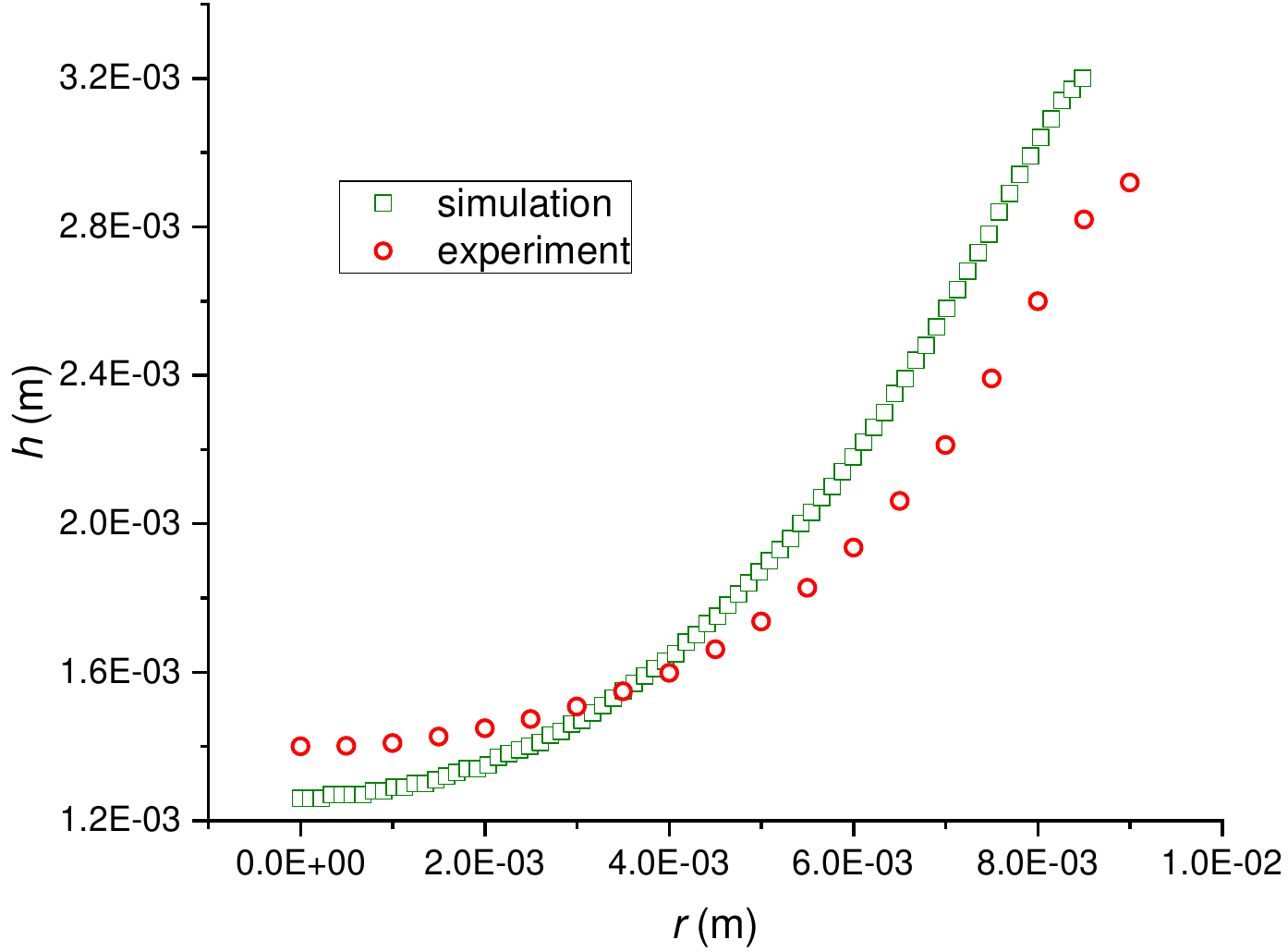}(b)\\
\includegraphics[width=0.4\textwidth]{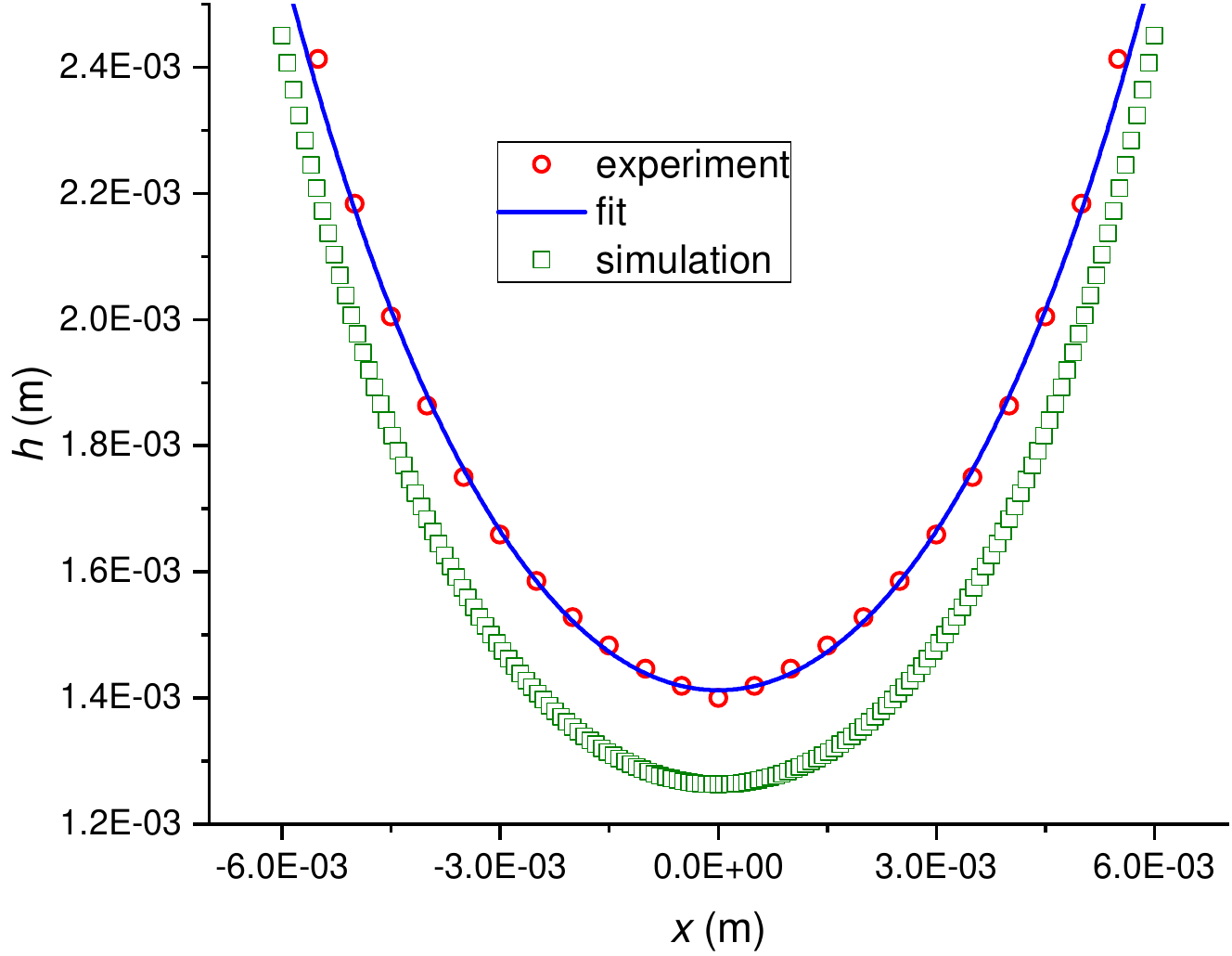}(c)
	\caption{Shape of the two-phase interface (``liquid--air'') along (a) the wall of square cell measured by analyzing optical images of the meniscus on the wall, (b) the square diagonal, and (c) its midline measured by the laser scanning method.}
	\label{fig:MeniscusProfileInSquareCell2D}
\end{figure}

Several cross-sections of meniscus profiles for a square cell are shown in Fig.~\ref{fig:MeniscusProfileInSquareCell2D}. For experimental data along the wall of the square cell, an approximation was constructed (Fig.~\ref{fig:MeniscusProfileInSquareCell2D}a). The resulting function was used as a boundary condition in numerical modeling (see details in Appendix~\ref{app:SquareCell}). The numerical results show qualitative agreement with experimental results (Figs.~\ref{fig:MeniscusProfileInSquareCell2D}b, \ref{fig:MeniscusProfileInSquareCell2D}c).

For the diagonal cross-section, the model yields slightly underestimated results in the central region compared to experiment (Fig.~\ref{fig:MeniscusProfileInSquareCell2D}b). Near the cell corner, we observe slightly overestimated results instead. The maximum error reaches about 14\% at point $r \approx$ 8.5~mm.

For the cross-section along the square's midline, the numerical solution shows slightly underestimated results relative to experiment (Fig.~\ref{fig:MeniscusProfileInSquareCell2D}c). The maximum error is approximately 10.7\% at point $x=0$.

\section{Conclusion}

This work presents a methodology for obtaining data on meniscus shapes in cells of various geometries, combining experimental measurements and mathematical modeling. The procedure begins by measuring the wetting meniscus height on the cell wall using optical image analysis and measuring the liquid layer thickness at the cell center via a contact method. These local measurements are then used with a relatively simple mathematical model to obtain complete meniscus shape data.

 An alternative approach could involve 2D interpolation of experimental data.  However, this method has several disadvantages. First, it requires measurements at significantly more points (at least an order of magnitude more), which would be time-consuming. Second, for each specific case (cell shape, material, liquid type, etc.), an appropriate interpolation function must be adjusted.

 The proposed approach uses 1D interpolation only to obtain the local thickness function $h$ along a specific cross-section or cell wall. Typically, a 4th--9th order polynomial suffices, and this doesn't require much time as the coefficients can be determined automatically using specialized software. The described mathematical model predicts meniscus shapes with an error of no more than 9--14\%, which may be acceptable for many applications.

 For applications requiring higher quantitative accuracy, the model can be refined by using the original curvature formula~\eqref{eq:SurfaceCurvatureFormula} instead of the simplified version~\eqref{eq:SurfaceCurvatureSimplified}, though this introduces nonlinearity and significantly increases the computational complexity. The work considers examples with cylindrical, square, and triangular cells, but we expect that this methodology can also be applied to other cell shapes, including those with more complex and asymmetric geometry. Only further research in this direction will allow this hypothesis to be confirmed or refuted.

\bmhead{Acknowledgements}

The work was supported by the Russian Science Foundation (Project No. 22-79-10216). The authors acknowledge D.~S.~Klyuev for assistance with experimental setup configuration, I.~V.~Vodolazskaya for discussions on the applicability of the Helmholtz equation in the mathematical model, and S.~A.~Kolegova for help with the translation of the text.

\section*{Declarations}
\bmhead{Funding}

Grant number 22-79-10216 from the Russian Science Foundation (\href{https://rscf.ru/en/project/22-79-10216/}{https://rscf.ru/en/project/22-79-10216/}).

\bmhead{Conflict of interest}

The authors declare no conflict of
interest.

\bmhead{Data Availability}

The data that support the findings of this study are available from the corresponding author upon reasonable request.

\bmhead{Author contribution}

KK: writing original draft, simulation, visualization, funding acquisition; VF: experimental set up, investigation, data collection;
NI: review and editing, experiment management, data analysis.

\begin{appendices}

\section{Square cell}\label{app:SquareCell}
For the case of a square cell, we assume that one of its sides is parallel to the $Ox$ axis and the other~--- to the $Oy$ axis. To determine the unknown constant, we consider the point $(x=0,\,y=0)$~--- the cell center. The meniscus profile shape in two mutually perpendicular cross-sections intersecting at this point coincides when considering either the square midlines or its diagonals. Taking this into account, from~\eqref{eq:HelmholtzEquation} for the central point we obtain the expression
\begin{equation}
	\left. \mathrm{C} =  \rho g f_0 - 2 \sigma f_{xx}\right|_{x=0},
\end{equation}
where $f$ is the approximation formula obtained using the least squares method (LSM) based on experimental data of the meniscus shape along the square midline (see Fig.~\ref{fig:MeniscusProfileInSquareCell2D}c),
$$
f(x) = \sum_{i=0}^4 k_i x^i.
$$
We use the notation $f$ instead of $h$ to emphasize that we are referring to the shape of the free surface along a specific cross-section, to whose direction the coordinate $x$ is tied. Here, $f_0$ denotes the function value at the central point, $f_0 = f(x=0)$. Using LSM in Origin 2019 software, the following coefficient values were determined: $k_0 = 0.141\times10^{-2}$, $k_1 = 0.2374841\times10^{-16}$, $k_2 = 26.74174$, $k_3 = -0.156871\times10^{-11}$, $k_4 = 150232.99512$. Thus, the value $\mathrm{C}\approx 6.3$ was obtained.

\section{Triangular cell}\label{app:TriangularCell}

Let us assume that one side of the triangle lies on the $OX$ axis, with one vertex located at $(x=0,\,y=0)$. Using experimental data on the meniscus shape along the wall, we construct an approximation formula via least squares method (LSM):
$$
f_\parallel (x) = \sum_{i=0}^9 l_i x^i,
$$
with the following coefficient values: $l_0 = 327\times10^{-5}$, $l_1 = -0.05527$, $l_2 = -82.19095$, $l_3 = 16130.99279$, $l_4 = -593806.82563$, $l_5 = -1.09081\times10^8$, $l_6 = 1.28383\times10^{10}$, $l_7 = -5.08505\times10^{11}$, $l_8 = 6.95934\times10^{12}$, and $l_9 = -2.90066\times10^{11}$ (Fig.~\ref{fig:filmShapeInTriangularCell2D}a).

To determine the constant $\mathrm{C}$, we use the point $(x=a/2,\,y=0)$ corresponding to the midpoint of the triangle side (side length $a\approx 18.2$~mm). Conceptually connecting this point to the opposite vertex with a line segment, we note that for an equilateral triangle this segment simultaneously represents the median, altitude, and angle bisector.

Considering a longitudinal cross-section coinciding with this segment and perpendicular to the $XY$ plane, we construct a meniscus shape approximation from experimental points using LSM along this cross-section:
$$
f_\perp (y) = \sum_{i=0}^4 m_i y^i,
$$
with coefficients: $m_0= 0.198\times 10^{-2}$, $m_1 = -0.13013$, $m_2 = 18.56667$, $m_3 = -1283.7233$, $m_4 = 56640.80432$ (Fig.~\ref{fig:filmShapeInTriangularCell2D}b).

From~\eqref{eq:HelmholtzEquation}, for the point $(x=a/2,\,y=0)$ we obtain:
\begin{multline*}
\mathrm{C} = \rho g f_\parallel (x=a/2) - \\
\sigma (\left. f_{\parallel xx} \right|_{x=a/2} + \left. f_{\perp yy}\right|_{y=0})\approx 15.25.
\end{multline*}

\section{Cylindrical cell}\label{app:CylindricalCell}

For the cylindrical cell case, it is convenient to use polar coordinates $(r,\,\varphi)$. Taking into account the axial symmetry of the meniscus shape, we exclude the angular coordinate $\varphi$ from consideration. The radial coordinate can be expressed in terms of Cartesian coordinates using the formula $r = (x^2 + y^2)^{1/2}$. We denote the local height of the liquid layer in the cylindrical cell as a function $H(r)$, from which we can transition to $h(x,y)$ if necessary, $$h(x,y) = H(\sqrt{x^2 + y^2}).$$

Equation~\eqref{eq:HelmholtzEquation} is rewritten as:
 \begin{equation}\label{eq:HelmholtzEquationCylindrical}
	- (r H^\prime)^\prime  + \rho g r H / \sigma = \mathrm{C} r / \sigma.
\end{equation}
Here, the notation $H^\prime$ denotes the derivative of the function $H$, $H^\prime = dH/dr$. Next, equation~\eqref{eq:HelmholtzEquationCylindrical} must be supplemented with two boundary conditions: $H(0)= H_0$ and $H^\prime(0) = 0$. The second condition follows from axial symmetry. The value $H_0$ is taken from experiment, $H_0 \approx 1.92\times 10^{-3}$ m.

To determine the constant $\mathrm{C}$, we integrate the left and right sides of equation~\eqref{eq:HelmholtzEquationCylindrical}, beforehand multiplying them by $2\pi$:
\begin{multline*}
  \frac{2 \pi \mathrm{C}}{\sigma} \int_0^R r \, dr + \\
  2 \pi \int_0^R (r H^\prime)^\prime \, dr - \frac{2 \pi \rho g}{\sigma} \int_0^R (r H) \, dr = 0.
\end{multline*}
Additionally, we account for the extra boundary condition $H^\prime (r=R) =\cot \theta$, where $\theta$ is the contact angle ($\theta \approx 1.34$ rad). Note that:
$$
2 \pi\int_0^R (r H) \, dr = V,
$$
where the liquid volume $V\approx 3\times 10^{-7}$ m$^3$, yielding:
$$
\mathrm{C} = \frac{\rho g V}{\pi R^2} - \frac{2 \sigma \cot \theta}{R} \approx 15.8.
$$

\begin{figure}[h]
	\centering
	\includegraphics[width=0.35\textwidth]{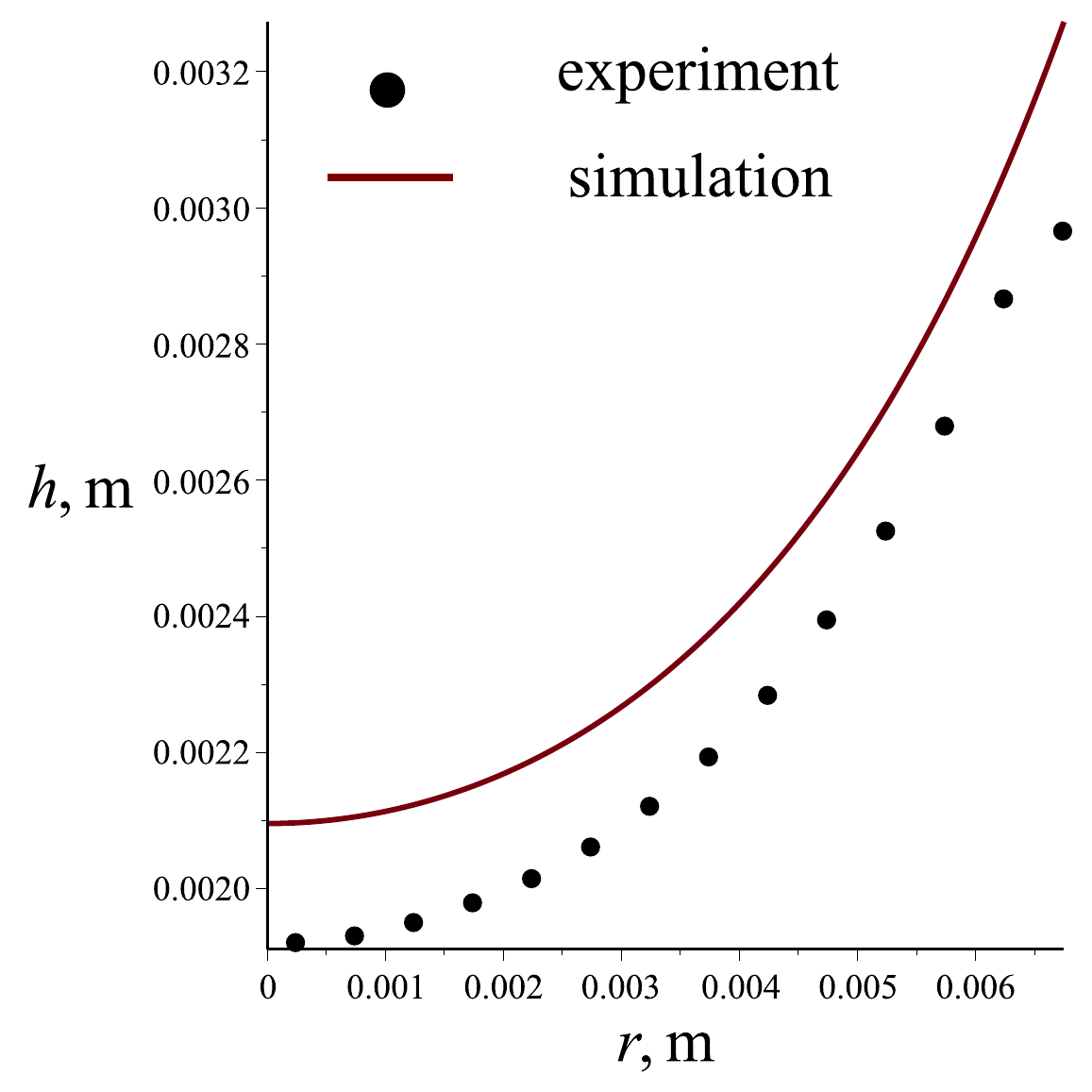}
	\caption{Results of the numerical calculation using the equivalent height $H_e$ and experimental measurements of the meniscus shape in a cylindrical cell.}
	\label{fig:h_experVsNumericCylinder_equivalentHeight}
\end{figure}

In the absence of experimental data for the value of $H_0$, it is sometimes acceptable to use the equivalent height $H_e = V/\left(\pi R^2 \right)$. This would be the height of the liquid layer if capillary forces were insignificant. The boundary condition then takes the form $H(0)= H_e$. In our case with a cylindrical cell, we obtain $H_e \approx$ $2.1\times 10^{-3}$~m. Note that using the equivalent height yields results that qualitatively agree with the experiment (Fig.~\ref{fig:h_experVsNumericCylinder_equivalentHeight}). The value is overestimated along the entire radius of the cell. The error does not exceed  10\%.

\end{appendices}

\bibliography{KolegovFliaginIvanova2025}
\end{document}